# Searching for Technosignatures: Implications of Detection and Non-Detection

**An Astro2020 Science White Paper**

**Thematic Areas:** Planetary Systems; Other

**Principal Author:**
Jacob Haqq-Misra
Blue Marble Space Institute of Science
jacob@bmsis.org
206-775-8787

**Co-authors:**

Anamaria Berea
University of Central Florida
anamaria.berea@ucf.edu
571-314-3768

Amedeo Balbi
University of Rome "Tor Vergata" (Italy)
balbi@roma2.infn.it
+39 06 7259 4565

Claudio Grimaldi
Ecole Polytechnique Fédérale de Lausanne
claudio.grimaldi@epfl.ch
+41 21 693 34 46

**Abstract**
The search for technosignatures from the Galaxy or the nearby universe raises two main questions: **What are the possible characteristics of technosignatures?** and **How can future searches be optimized to enhance the probability of detection?** Addressing these questions requires an interdisciplinary approach, combining i) the study of Anthropocene as a planetary transition and thus a possible proxy also for other planets, ii) the active search for technosignatures in the radio/optical and infrared range, and iii) the statistical modelling of technosignatures and Bayesian inference methods to learn from both detection and non-detection. **This strategy (combining modelling and observations) offers the best scientific opportunity in the next decade to discover the possible existence of technological civilizations beyond Earth.**

**Introduction**

Human influence on the biosphere has been evident at least since the development of widespread agriculture over 10,000 years ago, and some stratigraphers have suggested that the activities of modern civilization indicate a geological epoch transition from the Holocene to the Anthropocene. Our perspective as a civilization living within this transition allows us to contemplate the emergence of technological civilization in the context of planetary-scale processes (Grinspoon 2016). The Anthropocene may even represent a predictable planetary transition in general, to the extent that any energy-intensive species should drive changes in its biosphere (Frank et al. 2017). Examining the Anthropocene epoch through the lens of astrobiology can help to understand the future evolution of life on our planet and the possible evolution of technological, energy-intensive life elsewhere in the universe.

From an astrobiological perspective, drastic planetary changes such as the Great Oxygenation Event (or Oxygen Catastrophe) at the beginning of the Proterozoic, the Neoproterozoic Snowball Earth episodes, the Paleocene-Eocene Thermal Maximum (McInerney & Wing 2011), or the Permian-Triassic extinction event have led to major shifts in the dominant forms of life on Earth. Such events illustrate the ability of life to act as a transformative process on a planet, shaping the conditions that will accommodate future lifeforms.

Exoplanet atmospheric characterization is a related emerging area of interest that provides additional data for understanding how atmospheres evolve, which includes recent discoveries like Proxima Centauri b, the TRAPPIST-1 system, LHS1040b, and Ross 128b. Comparative modeling studies within the exoplanet science community could also benefit from interdisciplinary collaboration with the Earth climate community, especially to support mutual model development goals.

Besides natural causes for planetary-scale changes, any growing technological civilization living on a finite planet will face limits and consequences to growth, while enduring self-induced or extant threats that compound with time. The limits imposed by thermodynamics on a growing civilization suggest that large–scale effects such as the increase of the planetary equilibrium temperature due to enhanced energy consumption could be a universal feature of planets that have undergone an Anthropocene-like transition (Frank et al. 2017). The predicted fractional change in temperature ($\Delta T/T \sim 2\%$) at this thermodynamic limit corresponds to a world power use of $O(10^{16})$ W (compared to the $O(10^{13})$ W for today), or about 7% of the incoming solar radiation (Mullan & Haqq-Misra 2019). The $O(10^{16})$ W energy limit coincidentally corresponds to a "Type-I" civilization according to the Kardashev scale, which may point toward a fundamental limit of the observational imprint of a developing civilization. Frank et al. (2017) have even suggested a nominal classification scheme, which recognizes a series of evolutionary steps for a planetary system as geologic, biospheric, and technological processes emerge (see the figure on next page). This approach suggests that Earth today is between a "class IV" and "class V" planet, indicating the emergence of a global, and possibly detectable, technosphere.

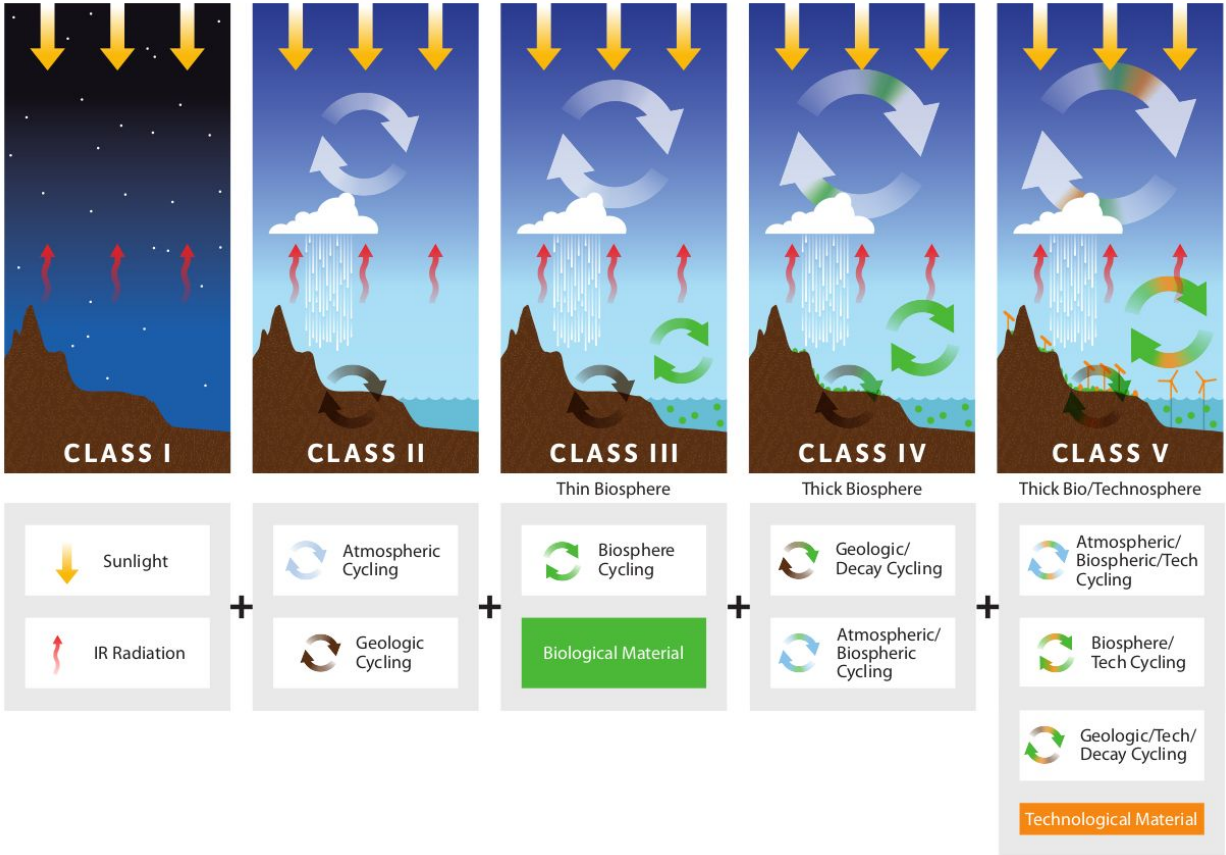

**Searching for Energy-Intensive Civilizations**

The global reach of human technology is evidence of the ongoing hybridization of Earth's biosphere with our energy-intensive civilization. The trajectory that led humanity to our present state, and the likely future outcomes of our experiments with exponential energy consumption, provides a plausible example of how to search for energy-intensive civilizations elsewhere. This does not necessarily imply that other inhabited planets will follow the same trajectory as life on Earth. Instead, the search for extraterrestrial intelligence (SETI) tends to operate with the working hypothesis that anything that happened here on Earth, or that is possible to happen in the future, remains a plausible option for guiding the search for other civilizations. (Note that in this white paper, "SETI" is presumed to be synonymous with the "search for technosignatures" of any kind and not necessarily limited to radio wavelengths.)

*Radio Technosignatures*

Perhaps the most well-known effort to detect technosignatures is the search for artificial radio and optical emissions (sometimes known as "radio and optical SETI"), which has been pursued since the 60's, albeit discontinuously. While the search has produced null results, the fraction of the multidimensional search space explored so far is so small that the non-detection to date has only a very modest informative value. The current state of ignorance has been recently quantified by Wright et al. (2018), who estimated that the fraction of the search space that has been explored is comparable to the ratio of the volume of a small swimming pool to that of the Earth's ocean. Even if the search space is restricted to emissions in the frequency range 1-2 GHz with an equivalent isotropic radiated power 100 times larger than that of the Arecibo radar, the probed fraction does not exceed $10^{-5}$ (Grimaldi & Marcy 2018).

While it is not possible at present to draw any sensible conclusion about the existence of non-natural EM emissions, let alone their abundance, there is the potential in the next decade to explore a larger fraction of the Galaxy over a broader range of wavelengths. This expectation is justified by recent large-scale search initiatives and the impressive progress in detection technologies, computing power, and big data mining techniques. In particular, the development of aperture arrays instruments and large-scale, wide-field telescope arrays offer significant advantages compared to single-dish radio telescopes, as larger portions of the sky can be searched in less time and with competitive sensitivities (Siemion et al. 2015, Garrett et al. 2017).

Radio SETI has so far managed to continue its efforts by appealing to the private sector for funding, such as the Breakthrough Listen initiative (Enriquez et al. 2017) as well as nightly surveys by the SETI Institute using the Allen Telescope Array (Harp et al. 2016). SETI represents an important objective of astrobiology, as progress in identifying and characterizing exoplanets also allows SETI to select better targets. This is an area for continued collaboration, in order to allow the observational and theoretical habitability studies from within astrobiology to also benefit SETI research (Frank & Sullivan 2016).

*Non-Radio Technosignatures*

Spectral signatures provide a way to characterize a planet's atmosphere, with a sizable astrobiology literature on possible atmospheric biosignatures that could indicate the presence of surface life (*e.g.,* Schwieterman et al. 2018). Spectral technosignatures are a particular spectral signature that would indicate the presence of a technological civilization on the planet (Schneider et al. 2010; Stevens et al. 2016). Examining the effects of energy-intensive human civilization on Earth's climate, both today and in likely future trajectories, can help to identify plausible technosignatures that might be observed with the next generation of space telescopes.

The terraforming of otherwise uninhabitable planets within a planetary system is an example of a possible technosignature, where powerful artificial greenhouse gases may be deployed to warm a planet outside the formal habitable zone (Fogg 2010). Such planets may be identified from the spectral features of greenhouse gases such as perfluorocarbons (PFCs), which are not known to otherwise occur in high abundances. Spectral technosignatures would produce the most observable features in the infrared portions of the electromagnetic spectrum, specifically in the thermal infrared window region at 8-12 μm for greenhouse gases. Conceptual studies of space telescopes capable of imaging terrestrial planets in the mid-infrared were previously studied, such as NASA's Terrestrial Planet Finder Infrared (TPF-I) space-based interferometer design concept (Beichman et al. 2006), and ESA's Darwin concept (Cockell et al.

2009), although neither is currently under consideration by either agency. The Origins Space Telescope (OST) concept is currently under study, which could resolve terrestrial planet features in the 8-12 μm range (Cooray et al. 2017).

The search for megastructures, such as Dyson swarms or other artifacts of extraterrestrial engineering, complements existing spectral surveys. The observation of anomalous absorption in the KIC 8462852 system (also known as "Boyajian's Star") prompted speculation on the possibility of detecting megastructures through transit photometry (Wright & Sigurdsson 2016; Gaidos 2017). Astrobiologists may therefore inevitably find themselves part of this discussion, particularly if future missions detect other unusual transit or spectral features.

**Inferring from detection or non-detection**

The data potentially gathered by future searches for technosignatures (either radio or non-radio) can have great informative value. The rationale behind this statement is that both the detection and non-detection are fundamental data in the process of inferring the existence and the possible population of technological civilizations in the Galaxy or in the nearby universe. Although such a deduction process would be a natural course of action in this field, it has not been systematically implemented so far. For example, the search for technosignatures in the radio spectrum has traditionally privileged technical aspects of the detection process at the expense of more theoretical approaches. The inferring process requires instead advanced statistical modeling to calculate the detection probability of technosignatures.

Whether we consider technosignatures originating from waste heat processes on other planets or from transmissions of radio/optical signals, the necessary condition for their detection is that the associated electromagnetic emissions (aka, the technosignature messengers) cross our planet. In addition to modelling the intrinsic properties of the emissions (i.e., their number density, degree of directionality, spectral properties), an effective theoretical framework should also consider scattering and attenuation effects of the interstellar medium as well as the instrumental characteristics of the detectors and data-processing software. In this way, the implications of non-detection/detection within a subset of the total search space can be drawn from a statistical Bayesian analysis, which is the most effective inference method when dealing with partial empirical knowledge. Specifically, statistical models of non-natural electromagnetic emissions allow the building of the likelihood functions associated to the event of detection or non-detection. These are then used via Bayesian analysis to infer the population and properties of the technosignatures in the entire Galaxy or in the local universe.

A first attempt to carry out such an analysis in this field has been described by Grimaldi & Marcy (2018), where fictitious searches were assumed to scan the entire sky for radio signals in the range 1-2 GHz. Depending on whether such fictitious searches detected a signal or not, the Bayesian analysis by Grimaldi & Marcy (2018) allowed to infer the posterior probability distribution of <k>, the configurational average of the number of signals crossing Earth emitted from the entire Galaxy. This kind of analysis can be extended from fictitious to real search programs in order to construct a probability landscape for the signals crossing Earth, which can be updated each time new data on a significant portion of the search space become available. If the non-detection turns out to be a persisting outcome of future searches, the Bayesian inference can inform in advance what would be the limit of the search space beyond which the probability of detection is small enough to call into question further searches. Alternatively, in the event that a detection is confirmed, depending on the extent of the search space in which the signal has been discovered, the new inferred distribution of detectable emissions may guide future searches to optimize the chances of further detections.

**Conclusion and Recommendations**

The study of the anthropocene as a geological epoch, and its implication for the future of civilizations, is an emerging transdisciplinary field in which astrobiology can play a leading role. In particular, it can inform about suitable models of both radio and non-radio technosignatures, which can be implemented in the formulation of Bayesian methods to infer the possible population of technosignatures from both the detection and non-detection. We recommend the following complementary approaches toward making significant progress in this area:

- The search for life would benefit from the development of missions such as LUVOIR, HabEx, and OST, which will provide the best opportunity in the coming decades to observe terrestrial biosignatures. Future decadal surveys should consider mission concepts similar to TPF-I and Darwin, or even a lunar observatory, in order to characterize biosignatures and possible technosignatures in the thermal infrared region.
- International collaborative efforts to search for technosignatures should be pursued, including traditional radio and optical SETI, spectral signatures in the 8-12 μm range, and other evidence of thermodynamic disequilibrium.
- The development of statistical, computational, and theoretical methods is a necessary condition to understand the character and detectability of plausible technosignatures in advance or in conjunction with the development of any mission capable of detecting such technosignatures.